\documentclass[sigconf]{acmart}
\usepackage{amsfonts}
\usepackage{multirow}
\usepackage{comment}
\usepackage{amsfonts}
\usepackage{makecell}
\usepackage{graphics}
\usepackage{subfigure}
\usepackage{stfloats}

\AtBeginDocument{%
  \providecommand\BibTeX{{%
    \normalfont B\kern-0.5em{\scshape i\kern-0.25em b}\kern-0.8em\TeX}}}

\copyrightyear{2023}
\acmYear{2023}
\setcopyright{acmlicensed}
\acmConference[ASIA CCS 2023]{ACM ASIA Conference on Computer and Communications Security}{July 10--14, 2023}{Melbourne, VIC, Australia}
\acmBooktitle{ACM ASIA Conference on Computer and Communications Security (ASIA CCS 2023), July 10--14, 2023, Melbourne, VIC, Australia}
\acmPrice{15.00}
\acmDOI{10.1145/3579856.3590339}
\acmISBN{979-8-4007-0098-9/23/07}

\begin{document}

\title{Masked Language Model Based Textual Adversarial Example Detection }

\author{Xiaomei Zhang}
\email{maruko1@email.swu.edu.cn}
\affiliation{%
  \institution{College of Computer and Information Science, Southwest University}
  \city{Chongqing}
  \country{China}
}

\author{Zhaoxi Zhang}
\email{zzx112358@email.swu.edu.cn}
\affiliation{
  \institution{School of Information Technology, Deakin University}
  \city{Geelong}
  \country{Australia}
}

\author{Qi Zhong}
\email{zhongq@deakin.edu.au}
\affiliation{%
 \institution{School of Information Technology, Deakin University}
 \city{Geelong}
 \country{Australia}
}

\author{Xufei Zheng}
\authornotemark[1]
\email{zxufei@swu.edu.cn}
\affiliation{%
  \institution{College of Computer and Information Science, Southwest University}
  \city{Chongqing}
  \country{China}
}

\author{Yanjun Zhang}
\email{Yanjun.Zhang@uts.edu.au}
\affiliation{%
  \institution{School of Computer Science, University of Technology Sydney}
  \city{Sydney}
  \country{Australia}  
}

\author{Shengshan Hu}
\email{hushengshan@hust.edu.cn}
\affiliation{%
  \institution{School of Cyber Science and Engineering, Huazhong University of Science and Technology}
  \city{Wuhan}
  \country{China}  
}

\author{Leo Yu Zhang}
\authornote{Xufei Zheng and Leo Yu Zhang are corresponding authors.}
\email{leo.zhang@griffith.edu.au}
\affiliation{%
  \institution{School of Information and Communication Technology, Griffith University}
  \city{Gold Coast}
  \country{Australia}
}

\renewcommand{\shortauthors}{Xiaomei Zhang et al.}

\begin{abstract}
Adversarial attacks are a serious threat to the reliable deployment of machine learning models in safety-critical applications. They can misguide current models to predict incorrectly by slightly modifying the inputs. 
Recently, substantial work has shown that adversarial examples tend to deviate from the underlying data manifold of normal examples, whereas pre-trained masked language models can fit the manifold of normal NLP data. 
To explore how to use the masked language model in adversarial detection, we propose a novel textual adversarial example detection method, namely Masked Language Model-based Detection (MLMD), which can produce clearly distinguishable signals between normal examples and adversarial examples by exploring the changes in manifolds induced by the masked language model. 
MLMD features a \textit{plug and play} usage (i.e., no need to retrain the victim model) for adversarial defense and it is agnostic to classification tasks, victim model's architectures, and to-be-defended attack methods.
We evaluate MLMD on various benchmark textual datasets, widely studied machine learning models, and state-of-the-art (SOTA) adversarial attacks (in total $3*4*4 = 48$ settings). 
{Experimental results show that MLMD can achieve strong performance, with detection accuracy up to 0.984, 0.967, and 0.901 on AG-NEWS, IMDB, and SST-2 datasets, respectively.} 
Additionally, MLMD is superior, or at least comparable to, the SOTA detection defenses in detection accuracy and F1 score. 
Among many defenses based on the off-manifold assumption of adversarial examples, this work offers a new angle for capturing the manifold change.
The code for this work is openly accessible at \url{https://github.com/mlmddetection/MLMDdetection}.
\end{abstract}



\begin{teaserfigure}
\end{teaserfigure}


\maketitle

\section{Introduction}
\label{Sec:Intro}
Machine learning (ML) has been widely used to solve natural language processing (NLP) tasks, such as sentiment analysis and machine translations, and has achieved remarkable performance. 
However, recent studies have shown that ML models are susceptible to adversarial examples, which are elaborately crafted by injecting small perturbations into normal examples \cite{Szegedy2014IntriguingPO, Goodfellow2015ExplainingAH, Alzantot2018GeneratingNL, Li2019TextBuggerGA}. 
These adversarial examples are imperceptible to human perception but can mislead the victim model to make incorrect predictions.
The existence of adversarial examples could significantly constrain the use of these ML models in security-sensitive applications, such as spam filtering and malware detection. Therefore, it is imperative to explore the fragility of victim models and then devise effective methods to thwart these attacks. 

A plethora of adversarial attack methods have been proposed, with many representative works focusing on constructing attacks for entire sentences \cite{Zhao2017GeneratingNA, Wang2020CATGenIR}. However, additional research has delved into the specifics of manipulating individual words and characters \cite{Ren2019GeneratingNL, Jin2020IsBR, Maheshwary2021GeneratingNL, Garg2020BAEBA, Li2020BERTATTACKAA, Li2019TextBuggerGA, Gao2018BlackBoxGO}.  
These attacks can generate adversarial examples that satisfy the lexical, grammatical, and semantic constraints of normal examples or are visually similar to the normal ones, making them difficult for humans to perceive. 
Recent studies \cite{Gilmer2018AdversarialS, Abusnaina2021AdversarialED, Shamir2021TheDM, Tanay2016ABT} have provided a new perspective to understand these adversarial examples, suggesting that they tend to leave the underlying data manifold of normal examples. 
The aforementioned adversarial attacks can thus be considered an attempt to push normal examples away from their manifold by carefully perturbing specific words or characters in a sentence. 

Meanwhile, in response to adversarial attacks, various defense techniques have been investigated. To force the victim model to be insensitive to suspicious inputs, adversarial training \cite{Goodfellow2015ExplainingAH, Zhou2020DefenseAA}, input randomization \cite{Wang2022RethinkingTA, Ye2020SAFERAS}, and synonym encoding \cite{Wang2021NaturalLA} have been widely studied. 
From the view of the data manifold, these defenses are effective because they can map the adversarial examples as close to the manifold of normal examples as possible, either by using adversarial examples as training data or processing inputs. Note that these defense methods require training from scratch \cite{Goodfellow2015ExplainingAH, Zhou2020DefenseAA, Wang2022RethinkingTA, Ye2020SAFERAS} or even modifying the model architecture \cite{Wang2021NaturalLA}.

\begin{figure*}[t!]
    \centering 
    \includegraphics[width=16cm]{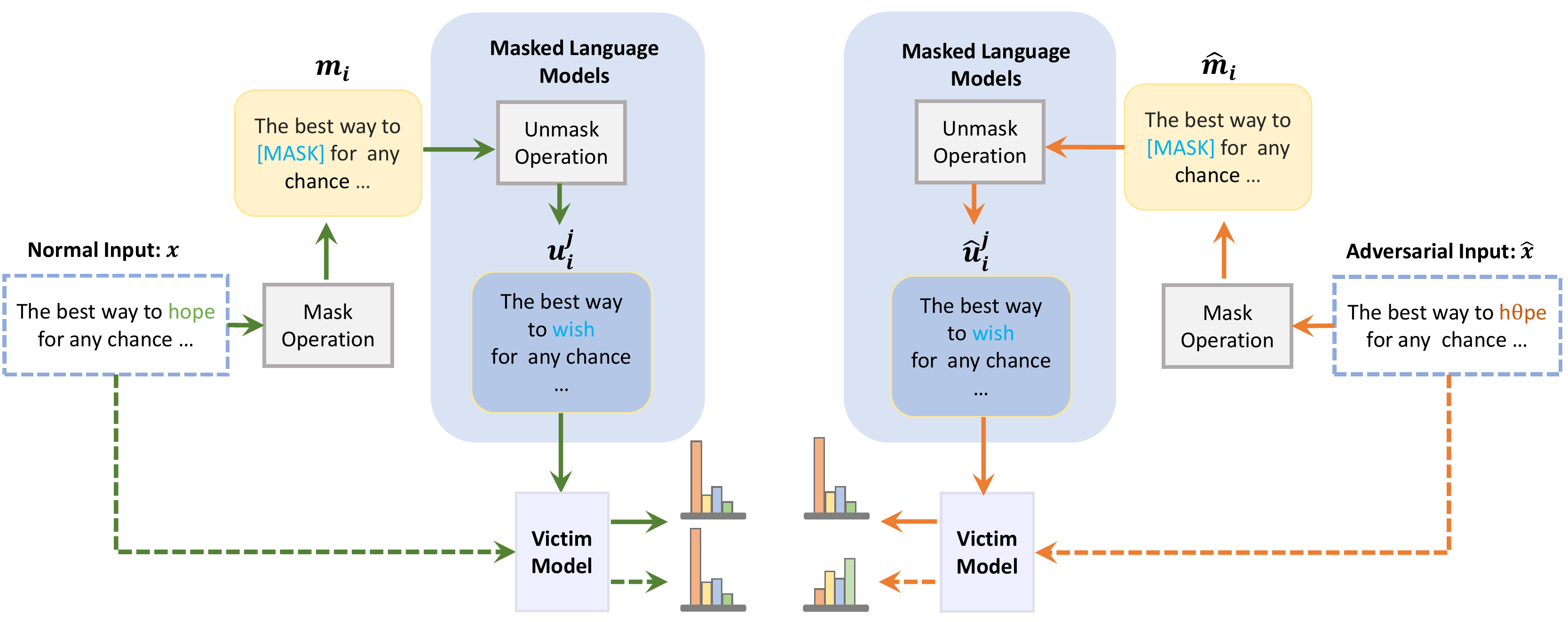}
    \caption{An overview of MLMD.} 
    \label{Fig:mlmdOverview}
\end{figure*}

As another branch of defense, detecting textual adversarial examples is relatively underexplored. The very first adversarial detector \cite{Zhou2019LearningTD} employed a discriminator network to identify suspicious tokens and an estimator network to rectify them. 
It is worth noting that adversarial detection has several advantages over the strategy that strengthens the robustness of the victim model.

Specifically, an adversarial detector can be easily deployed as a top-up ~\emph{plug and play} module, without affecting the performance of the victim model. 
In contrast, traditional defense techniques such as adversarial training and input randomization may have some (more or less) impact on the model's prediction capability.
In addition, an adversarial detector also features its \emph{online} property, meaning that it can be continuously used to detect adversarial examples as the victim model evolves with different datasets or model architectures.

To detect adversarial text examples, several studies have attempted to trigger changes in data manifold by substituting some input tokens with their synonyms or special tokens \cite{Mosca2022ThatIA, Wang2022DetectingTA, Moon2022GradMaskGT}. 
However, these examples with substituted tokens cannot always be mapped to the manifold of normal examples, thus making the observed manifold change less useful for adversarial detection. 
Concurrently, extensive work has indicated that pre-trained masked language models could capture information about the manifold of normal examples after accomplishing a Masked Language Modeling (MLM) objective on large-scale normal data \cite{Ng2020SSMBASM, Hendrycks2020PretrainedTI, Xu2021UnsupervisedOD}.  
The MLM objective is composed of two operations.  
The first is the mask operation, which means a percentage of the tokens will be masked out in the example to stochastically perturb the example of the current manifold. 
Then an unmask operation projects it back on by training the model to predict the masked tokens from the rest of the sentence. 
These findings inspire us to explore the potential of applying the masked language model and the MLM objective for {textual} adversarial detection.

This paper proposes a novel textual adversarial example detection method called Masked Language Model-based Detection (MLMD), as shown in Fig.~\ref{Fig:mlmdOverview}.
MLMD utilizes a pre-trained masked language model to distinguish between normal examples and adversarial examples by exploring changes in the manifold. 
For an adversarial example, MLMD pushes it away from its original manifold with the mask operation, but the subsequent unmask operation projects the masked example onto the manifold which normal examples concentrate on. For a normal example, since its manifold fits well with the MLM objective of the masked language model, executing the mask and unmask operations of the MLM objective does not result in any change to the manifold. 
Unlike other defenses that exploit the change in the manifold, the key advantage of MLMD lies in the  MLM objective, which ensures a good fit of the manifold of normal examples.

Through a comparative analysis with 2 state-of-the-art (SOTA) detections \cite{Mozes2021FrequencyGuidedWS, Mosca2022ThatIA}, empirical results over 3 datasets (i.e., AG-NEWS, IMDB and SST-2), 4 victim model architectures (i.e., CNN, LSTM, BERT, and ALBERT) and 4 SOTA attack methods  (i.e., PWWS, TextFooler, DeepWordBug, and TextBugger) demonstrate that MLMD is more effective in detecting textual adversarial examples. 
Moreover, we also investigate the influence of selecting different (or removing) masked language models (e.g., BERT, ALBERT, or RoBERTa) for detection, using different masking rates, and fine-tuning the masked language model to better fit the manifold of the training data. 
These experiments demonstrate that MLMD's performance is robust to variations in these parameters and that fine-tuning the masked language model can lead to further improvements in detection performance. The main contributions of this work are summarized as follows:
\begin{itemize}
    \item We unveil the power of masked language models for textual adversarial example detection  by exploiting the change in manifold through the mask and unmask procedures.

    \item We propose MLMD as a defense mechanism against adversarial examples, which is independent of the attack method or the victim model. Moreover, it does not require any retraining.

    \item We conduct extensive experiments using both traditional classifiers and transformer-based models for text classification tasks, evaluating MLMD's performance against several representative textual adversarial attacks. 
    Empirical results show that MLMD achieves comparable or even better performance compared to state-of-the-art detection algorithms. 
\end{itemize}

\section{Background and Related Works}

\subsection{Textual Adversarial Attacks}
Textual adversarial attacks aim to generate a specifically perturbed textual input, dubbed textual adversarial example, that is close to its corresponding raw textual input but can fool the victim model into making incorrect predictions.  
Specifically, besides being capable of fooling the victim, a qualified textual adversarial example should also meet three key properties \cite{Jin2020IsBR}: 
(1) human prediction consistency—the adversarial example should have the same human prediction as the raw; 
(2) semantic similarity—the adversarial example should have the same meaning as the raw, as judged by humans; 
and (3) linguistic fluency—generated adversarial examples should appear natural and grammatical.

According to the granularity of perturbation, textual adversarial attacks can be roughly divided into character-, word-, and sentence-level attacks. Sentence-level attacks take the whole raw textual input as the target of perturbation, and they are typically launched by
rephrasing important words or paragraphs  \cite{Iyyer2018AdversarialEG}, or using GAN \cite{Goodfellow2014GenerativeAN} to directly generate perturbations \cite{Zhao2017GeneratingNA, Wang2020CATGenIR}. However, the perturbation in sentence-level attacks is generally large and can be perceived by humans. For this reason, this study focuses on word- and character-level attacks, as in previous works \cite{Ren2019GeneratingNL, Jin2020IsBR, Li2019TextBuggerGA, Gao2018BlackBoxGO}.

\textbf{Word-level Attacks}: Word-level attacks have become mainstream in the research community due to their powerful attack performance and capability in maintaining semantic similarity and grammatical correctness.
These attacks target words in the raw textual input, and the most common approach is substitution-based.

To guarantee the attacks that are hard to be perceived by humans, some studies \cite{Ren2019GeneratingNL, Jin2020IsBR, Maheshwary2021ASB} proposed to generate small perturbations by selecting critical words and using synonym replacements. These synonyms are generated from the well-built network WordNet or selected from candidate words with similar embeddings. 
Inspired by the process of natural selection,  \cite{Alzantot2018GeneratingNL, Maheshwary2021GeneratingNL} used population-based optimization algorithms to search for the best perturbation while ensuring semantic similarity. 
However, direct synonym replacement may not be effective in preserving semantic consistency and linguistic fluency, many works~\cite{Garg2020BAEBA, Li2020BERTATTACKAA, Li2021ContextualizedPF} introduced language models as automatic perturbation generators.

\textbf{Character-level Attacks}: Character-level attacks typically design typos for numbers, letters, and special symbols in the raw textual input. Even though the textual adversarial example computed in this way causes a certain degree of semantic change, it does not affect human judgments because of its visual similarity to the raw textual input. 
HotFlip~\cite{Ebrahimi2018HotFlipWA} is an earlier attack method that perturbs characters based on an atomic flip operation. TextBugger~\cite{Li2019TextBuggerGA} and DeepWordBug~\cite{Gao2018BlackBoxGO}, similar to some mainstream word-level attacks, generate textual adversarial examples by identifying important words in an input {and disturbing characters of the identified words in an appropriate way}. VIPER \cite{Eger2019TextPL} randomly replaces characters with their visual nearest neighbors in a visual embedding space. 

From the perspective of manifold, these textual adversarial attacks elaborately design perturbations to push examples off the manifold of normal data. When the perturbation is strong enough, such as replacing enough characters or words that contribute the most to the victim's classification, the perturbed example can cross the decision boundary and cause the victim to make an erroneous prediction.
\subsection{Defenses for Textual Adversarial Attacks}
One of the most imperative purposes of exploring adversarial examples is to boost the robustness of the victim model. 
To mitigate the threat of adversarial attacks, researchers have proposed various defense methods, which can be bifurcated into two broad categories: robust prediction and adversarial detection.

\textbf{Robust Prediction}: This kind of defense aims to train a robust model that is immune to manipulated textual inputs. 
The most famous method for achieving this is adversarial training. It iteratively solves a two-layered $\min\max$ optimization problem: identifying adversarial examples against the current model to enrich the training set, and updating the model by minimizing the loss on the enriched training set~\cite{Goodfellow2015ExplainingAH, Zhou2020DefenseAA, Dong2021TowardsRA}. 
Though this idea is promising, {training two-layered optimization problems of this type can be very difficult to converge,} and the defensive approach may not be applicable to deployed models.
Empirical studies~\cite{Wang2022RethinkingTA, Ye2020SAFERAS} have found that randomization of inputs can invalidate a large portion of textual adversarial examples. 
Several defense methods \cite{Wang2021NaturalLA}, instead, aim to map similar inputs to similar encodings and train the victim model to eliminate possible adversarial perturbations.

It is also a promising direction to enhance the robustness of the victim from the perspective of data manifolds~\cite{Zhang2021SelfSupervisedAE}. A typical example is the Textual Manifold-based Defense \cite{Nguyen2022TextualMD}, in which the authors trained a generative model InfoGAN \cite{Chen2016InfoGANIR} on continuous representations to approximate the manifold of normal data and then projected inputs onto the learned manifold before classification. 
From the manifold perspective, adversarial training methods attempt to regulate the examples that are scattered at the manifold boundary to force the victim to make correct predictions. 
Randomization-based and encoding-based defense techniques map adversarial examples back to the manifold where the normal examples are located to eliminate adversarialness.

\textbf{Adversarial Detection}:
The goal of adversarial detection is to identify whether a textual input is maliciously perturbed and then reject the detected adversarial examples or eliminate the potential adversarial perturbations.
Typical examples are the defenses against the character-level attacks presented in \cite{Pruthi2019CombatingAM, Alshemali2019TowardMA}, where the authors designed a spell-checking system for detecting  spelling and syntax errors and correcting them before classification.
{For word-level attacks}, Mozes et al. presented Frequency-Guided Word Substitutions (FGWS) \cite{Mozes2021FrequencyGuidedWS} to identify textual adversarial examples based on the assumption that adversarial attack algorithms prefer exploiting words rarely exposed in the training set of the victim model. 
Recently, some works \cite{Mosca2022ThatIA, Wang2022DetectingTA, Moon2022GradMaskGT} have focused on eliminating adversarial perturbations by substituting words with their synonyms or special tokens and then detecting suspicious input by estimating the difference between the original input and the manipulated example.

Taking the off-manifold conjecture (discussed next in Sec.~\ref{Sec:Conjecture}) of adversarial examples into consideration, a detector merely needs to distinguish between off-manifold and on-manifold (i.e., normal) examples. From the perspective of manifold, both spell-checking (to resist character-level attacks) \cite{Pruthi2019CombatingAM} and substituting (key) words with synonyms or special tokens (to resist word-level attacks) \cite{Mozes2021FrequencyGuidedWS, Mosca2022ThatIA, Wang2022DetectingTA, Moon2022GradMaskGT} can be considered a way of promoting manifold change. Such change plays the role of a feature for building the detector. However, examples with substituted tokens or that have been spelling-checked will not necessarily be mapped to the manifold of normal examples as we mentioned in Sec.~\ref{Sec:Intro}. From this sense, a projection method that fits the normal data manifold better will naturally lead to better features for building a desired adversarial detector.

\subsection{Masked Language Models and The Off-manifold Conjecture}
\label{Sec:Conjecture}
As mentioned earlier, masked language models are pre-trained on large-scale unlabeled normal data with the MLM objective. This pre-training equips the models with the capacity of reconstructing input data. 
As the most representative example of the masked language model, 
BERT \cite{Devlin2019BERTPO} is a bidirectional language model trained with {the} MLM objective to learn general language representations from unlabeled data, leading to a series of breakthroughs in NLP. 
ALBERT \cite{Lan2020ALBERTAL} is pre-trained on the same data as BERT, but it adopts the n-gram masking strategy of the MLM objective. The MLM target consists of up to an n-gram of complete words, which encourages the model to capture more comprehensive language representations.
RoBERTa \cite{Liu2019RoBERTaAR} improves BERT by employing the dynamic masking strategy, which dynamically changes the masking pattern applied to training data. Additionally, it is trained over a variety of corpora with more sophisticated techniques, such as training the model longer and with bigger batches. Thus, it has the potential to better approximate the manifold of normal data. 

The off-manifold conjecture \cite{Shamir2021TheDM, Gilmer2018AdversarialS, Tanay2016ABT, Abusnaina2021AdversarialED} states that adversarial examples tend to deviate from the manifold of normal examples. 
This conjecture offers a new perspective to interpret the existence of adversarial examples and has attracted significant research attention \cite{Meng2017MagNetAT, Zhang2021SelfSupervisedAE, Yang2021ClassDisentanglementAA, Nguyen2022TextualMD}. 
In computer vision tasks, under the off-manifold conjecture, substantial detection approaches \cite{Meng2017MagNetAT, Zhang2021SelfSupervisedAE, Yang2021ClassDisentanglementAA} have approximated the manifold  of normal images by generative models and designed effective metrics to measure changes in the manifold to detect adversarial images. 
In NLP, Nguyen et al. \cite{Nguyen2022TextualMD} found that adversarial texts also tend to diverge from the manifold of normal ones. They defended the victim model by using the InfoGAN to characterize the manifold of normal texts and then projecting the input onto the approximated manifold.
Concurrently, many related studies \cite{Hendrycks2020PretrainedTI, Ng2020SSMBASM, Xu2021UnsupervisedOD} have shown that masked language models have the potential to fit the manifold of normal textual examples and can be used to improve out-of-distribution robustness, resulting in performance boosts. 
These findings inspire us to investigate a new line of textual adversarial detection. Specifically, relying on the property that masked language models can fit the manifold (of normal examples), we hypothesize that the behavior of off-manifold (i.e., adversarial) examples will differ from normal ones when projected by masked language models.

\section{Method}
\subsection{Notations}
In this paper, we mainly focus on detecting textual adversarial examples that aim to attack standard classification tasks. 
Let $f(\cdot)$ denote a victim model trained over pairs of an input sequence $x=\{w_{\textnormal{1}}, w_{\textnormal{2}}, \cdots, w_{n}\}\in \mathcal{X}$ of $n$ words and its ground-truth label 
$ y^* \in \{1, 2, \cdots, c\}$
with $c$ being the number of classes. 
During inference, given a test example $x_t$, the victim model $f$ outputs a confidence score vector $f(x_t)$, where $\sum_{y=1}^c f(x_t)_y = 1$. 
We denote the final prediction as $z(x_t)=\underset{y}{\arg\max}\, f(x_t)_y$.

An input sequence $x$ and its corresponding adversarial counterparts $\hat{x}$
should be either perceptually indistinguishable or semantically consistent, i.e., it satisfies 
$\hat{x} = x+\delta$, $z(\hat{x})\neq z(x)$, and $\left\| \delta \right\| <  \epsilon$.
Here, $\delta$ is known as the adversarial perturbation, which is bounded by the threshold $\epsilon$. 
To preserve grammatical correctness, semantic consistency, or perceptual unnoticeability, the perturbation $\delta$ is typically achieved through addition, removal, or substitution operations on the words or characters of the raw input text $x$.

\subsection{General MLMD Framework}
In this section, we describe our proposed method, MLMD, which consists of three parts: 
\begin{itemize}
    \item  A mask function $F_m$ that intentionally moves the input text off from its original manifold by corrupting raw input;
    \item An unmask function $F_u$ that uses the masked language model $\Phi$ to project the corrupted texts generated by $F_m$ back to the manifold of normal examples;
    \item A classifier $C_a$ that detects adversarial examples by capturing the difference induced by the change of manifold.
\end{itemize}
Formally, the MLMD detector is a distinguisher $\mathrm{d}(F_m, F_u, C_a)$: $\mathcal{X}\cup \hat{\mathcal{X}}\rightarrow\mathcal{Y}$, where $\mathcal{X}$ is the entire space of all normal texts, $\hat{\mathcal{X}}$ is the space of all adversarial texts, and $\mathcal{Y}=\{0, 1\}$ is the set of ground truth binary labels (with $1$ denoting adversarial texts). Since it is intractable to capture the entire space of {normal} and adversarial examples, we access it by estimating $\mathcal{X}\cup\hat{\mathcal{X}}$ with the dataset
$\mathcal{D}= \mathcal{D}^n \cup \mathcal{D}^a $, $x\in\mathcal{D}^n$ and $\hat{x}\in\mathcal{D}^a$.

\subsection{Mask and Unmask Procedures}

Given any input text sequence $x$ (either normal or adversarial), we map it to the mask manifold by feeding it into the mask function $F_m$ to get an ensemble of masked sequences as
\begin{equation}
    M=F_m(x,r), 
\label{Eq:masking}
\end{equation}
where $M=\{m_i|i\in[1, r*n]\}$, $m_i$ denotes $i$-th masked text generated by replacing the selected token $w_i$ in original input $x$ with [MASK], $r$ denotes the masking rate, and $n$ denotes the length\footnote{With loss of generality, we consider $r*n$ to be an integer.} of $x$. 
{The implementation of $F_m$ is not limited, including randomly masking words of $x$, or masking only key words of $x$.}

{Afterwards, a masked language model $\Phi$ is used to backfill the masked words for each sequence in the ensemble $M$.}
For each $m_i$ (i.e., the masked version of $x$), we retain the top-$k$ candidates restored by $\Phi$. This procedure is represented as:   
\begin{equation}
    U=F_u(M, k, \Phi),   
    \label{Eq:unMask}
\end{equation}
where $U=\{u_i^j| i\in[1, r*n], j\in [1,k]\}$ indicates the set of reconstructed text sequences. 
For a given $i$, $u_i^j$  represents the $j$-th ($j \in [1, k]$) backfilled candidate when unmasking the sequence $m_i$. 

Recall the discussions in Sec.~\ref{Sec:Conjecture}, if $x$ is a  normal example, then there is no change in the manifold through {the mask and unmask procedures}. This contrasts with the off-manifold to on-manifold movement when $x$ is adversarial.

\subsection{Adversarial Classifiers}
The mask and unmask procedures endow the normal and adversarial examples with significantly distinguishable signals. 
We now explore how to apply these signals to adversarial example detection. 

\subsubsection{Building Threshold-based Classifier}
For an input $x$, we define a distinguishable score $S_t(x, f, \Phi)$ for $x$ based on the masked language model $\Phi$ and {the} victim model $f$. 
Specifically, the score $S_t(x, f, \Phi)$ 
is calculated by
\begin{equation}
    S_t(x, f, \Phi)= \frac{ \sum\limits_{i=1}^{r*n}\sum\limits_{j=1}^{k}\mathbb{I}(z(x), z(u_i^j))}{r*n*k},
\label{Eq:disScores}
\end{equation}
where $z(x)=\underset{y}{\arg\max}\, f(x)_y$, $u_i^j \in U$  (defined by Eq.~(\ref{Eq:unMask})) is one reconstruction result by $\Phi$,
and
{$\mathbb{I}(\cdot, \cdot)$}
is the indicator function that returns 1 when the two operands equal each other.

Clearly, $S_t(x, f, \Phi)$ falls into $[0, 1]$. Based on the manifold conjecture discussed in Sec.~\ref{Sec:Conjecture}, $S_t(x, f, \Phi)$ tends to be large if $x$ is on-manifold (i.e.,  normal examples) and tends to be small if $x$ is off-manifold (see Fig.~\ref{Fig:STscores} for a visual validation). After obtaining the score $S_t(x, f, \Phi)$ over the dataset $\mathcal{D}$, the desired classifier $C_a$ can be easily obtained by setting an appropriate threshold $\tau$. {We empirically determine $\tau$ with a \textit{one-time} offline process by selecting the value that maximizes the detection accuracy and use the same $\tau$ for online adversarial example detection later on.}

\subsubsection{Building Model-based Classifier}
\label{subsubsec:ModelbasedDetector}
The threshold-based classifier described above only captures the changes in the label of raw input $x$ and the corresponding reconstructed texts contained in $U$. 
However, one-hot encoded labels discard important information about the decision boundary of the victim $f$, and may not be optimal for many security tasks, including defending against adversarial examples and membership inference attacks \cite{Wang2022DetectingTA, Carlini2021MembershipIA, Zhang2022EvaluatingMI}. 
As an alternative method, we propose an adversarial classifier that directly uses 
the confidence score $f(x)$ in the following. 

\textbf{Feature Engineering.} For an input $x$, we will have $r*n*k$ backfilled candidates $u_i^j$ after the mask and unmask procedures. {Inspired by the work in \cite{Mosca2022ThatIA}, we calculate a feature vector $FE = [fe_l]_{l=1}^{rnk}$ for $x$ as}
\begin{equation}
    fe_l= f(u_i^j)_{y^*} - \underset{y \neq y^*}{\max}f(u_i^j)_y, 
\label{Eq:featureVector}
\end{equation}
where $y^* = \underset{y}{\arg\max}\, f(x)_y$, $i \in [1, r*n]$, $j \in [1, k]$ and $l = (i-1)*k + j$. {Clearly, if $x$ is a normal example, $fe_l$ tends to be positive; otherwise, $fe_l$ tends to be negative.} We also sort the feature vector $FE$ in ascending order and denote the sorted feature vector as $\overline{FE}$.
By comparing the detection performance of the original and sorted feature vectors, we attempt to discuss the impact of the order of elements in the feature vector on detection in Sec.~\ref{Sec:model_based_performance}.

\textbf{Dataset and Binary Classifiers.} With the features available, we further construct the dataset as $\Gamma = \{ (FE_x, 0)\} \cup \{(FE_{\hat{x}}, 1)\} $ and $\overline{\Gamma} = \{ (\overline{FE}_x, 0)\} \cup \{(\overline{FE}_{\hat{x}}, 1)\}$
for $x \in\mathcal{D}^n$ and $\hat{x} \in\mathcal{D}^a$. 
Since the samples in either $\Gamma$ or $\overline{\Gamma}$ should be clearly separable if the manifold assumption holds, 
the architecture of the classifier is not expected to have a significant impact on the detection performance. As such, we train two different models, a multi-layer perceptron (MLP) model, and an XGBoost model, as the binary classifier $C_a$.  
To ensure consistent input dimension for the training of $C_a$ ($n$ may vary across different examples), 
we either pad the example with zeros or truncate it to the appropriate length.

\section{Experimental Setup}
\subsection{Datasets}
\label{Subsec:datasets}
We evaluate the detection capability of MLMD on three prevalent classification benchmark datasets: AG-NEWS \cite{Zhang2015CharacterlevelCN}, Internet Movie Database (IMDB) \cite{Maas2011LearningWV} and Binary Stanford Sentiment Treebank (SST) \cite{Socher2013RecursiveDM}, whose details are elaborated as follows. 
\begin{itemize}
    \item AG-NEWS: It is a dataset consisting of news articles divided into four topics: World, Sports, Business, and Sci/Tech. Its training set has 120K news articles, and its test set has 7.6K articles. The average length of an article is 43 words.
    \item IMDB: It consists of 50K movie reviews for binary sentiment (positive or negative) classification. 25K reviews are used as the training set and 25K reviews as the test set. The average length of a review in IMDB is 215 words.
    \item SST-2: SST is a corpus with fully labeled parse trees used to analyze the compositional effects of sentiment. Following the setting in \cite{Moon2022GradMaskGT, Mozes2021FrequencyGuidedWS}, we convert it into a binary dataset (SST-2) labeled with positive or negative sentiment labels. {This dataset incorporates a training set with 67K texts, a validation set with 0.8K sequences, and a test set with 1.8K examples. The average length of a text example is 20 words}. 
\end{itemize}



\subsection{Victim Models}
\label{subsec:vicmodels}
We test MLMD on a Convolutional Neural Network (CNN) 
\cite{Zhang2015CharacterlevelCN} and a Long Short-Term Memory (LSTM) \cite{Hochreiter1997LongSM}, which are considered as the backbone in many tasks. 
We also evaluate our method on two Transformer-based models BERT \cite{Devlin2019BERTPO} and ALBERT \cite{Lan2020ALBERTAL}, which are initialized with pre-trained weights and subsequently fine-tuned using downstream labeled data. 
All the aforementioned victim models are available in the TextAttack library \cite{Morris2020TextAttackAF} and have already been trained on the datasets mentioned in Sec.~\ref{Subsec:datasets}. 

\begin{figure*}[t]
    \centering 
        \subfigure[PWWS]{\includegraphics[width=4.35cm]{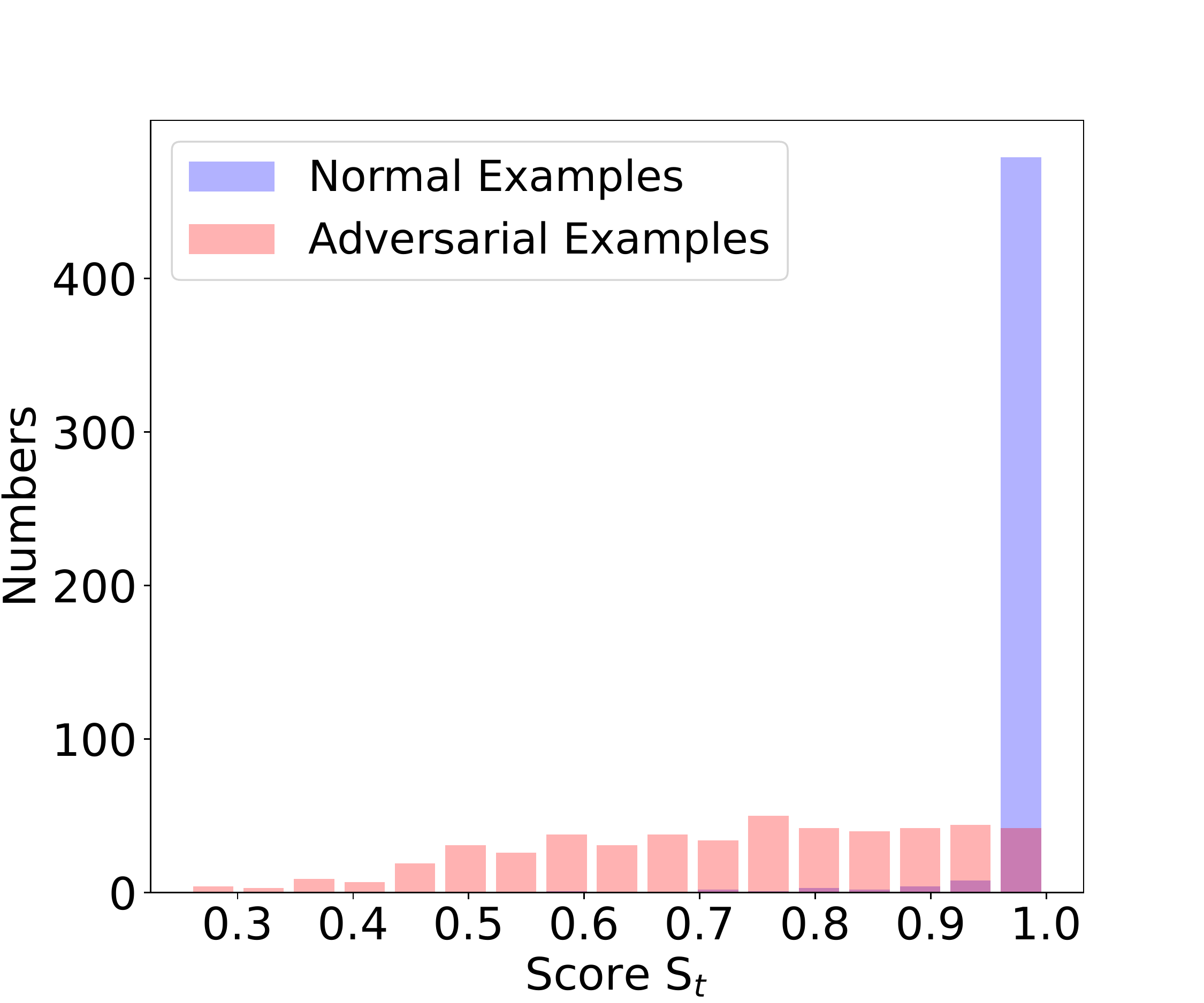}}
        \subfigure[TextFooler]{\includegraphics[width=4.35cm]{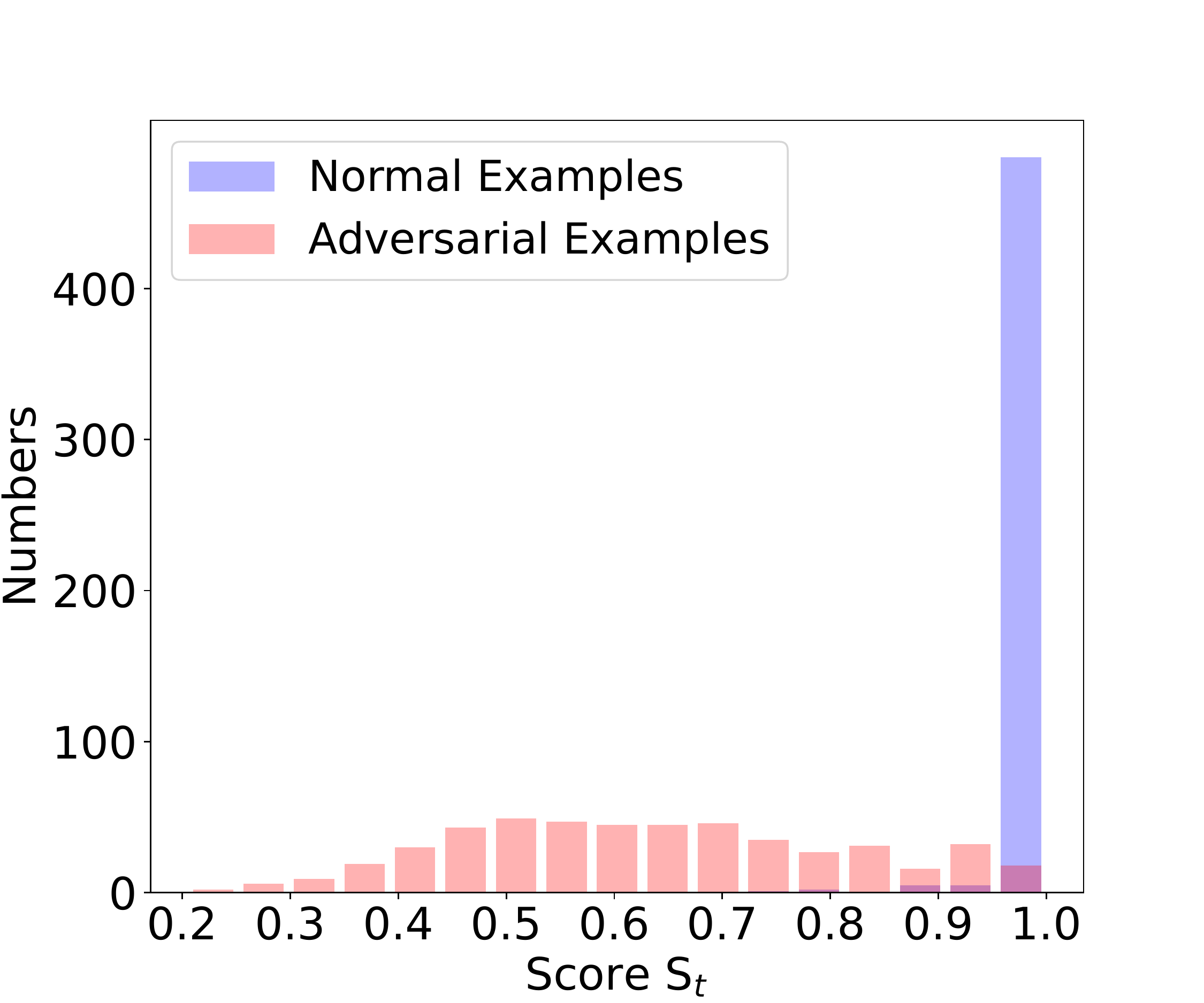}} 
        \subfigure[TextBugger]{\includegraphics[width=4.35cm]{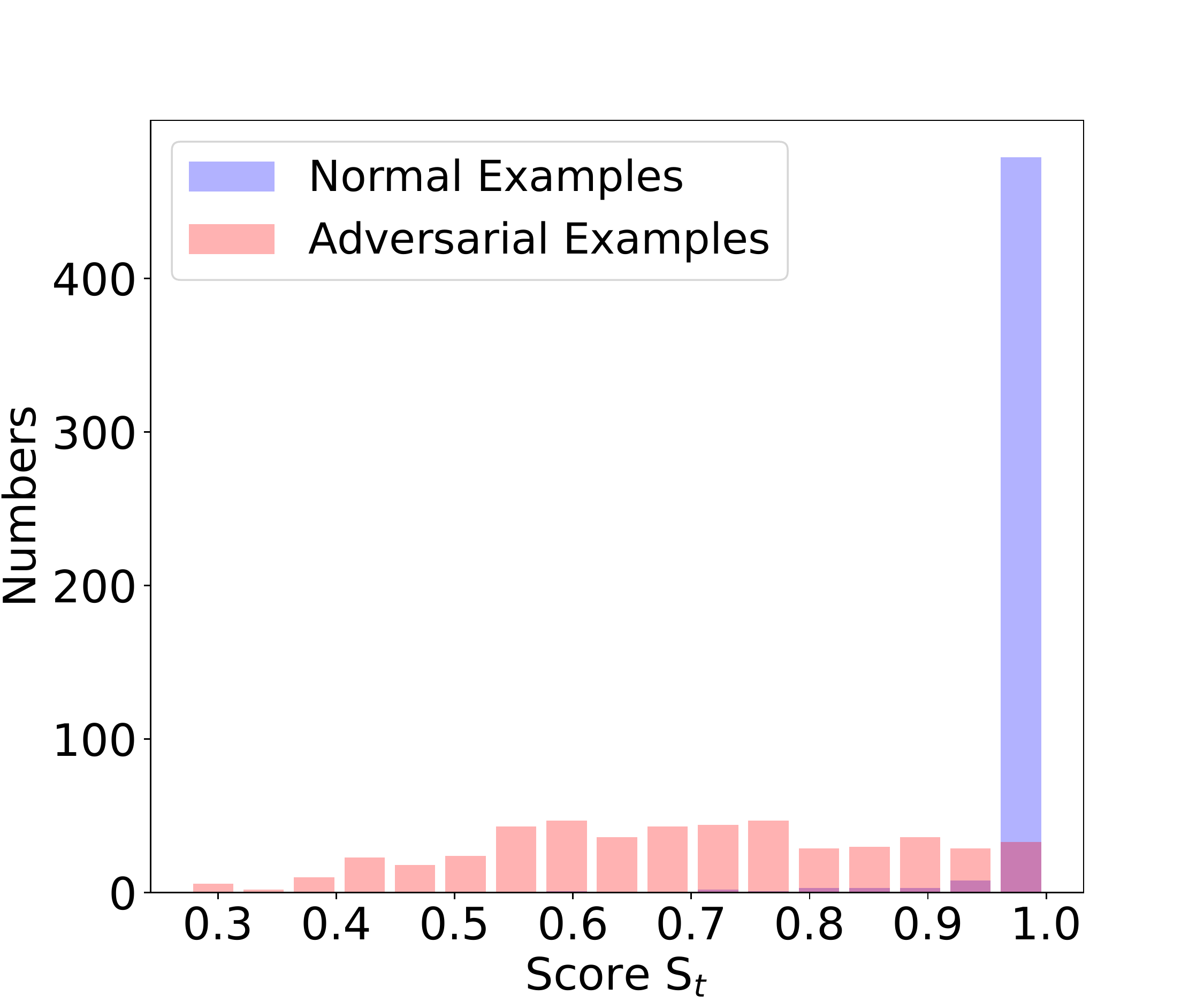}}
        \subfigure[DeepWordBug]{\includegraphics[width=4.35cm]{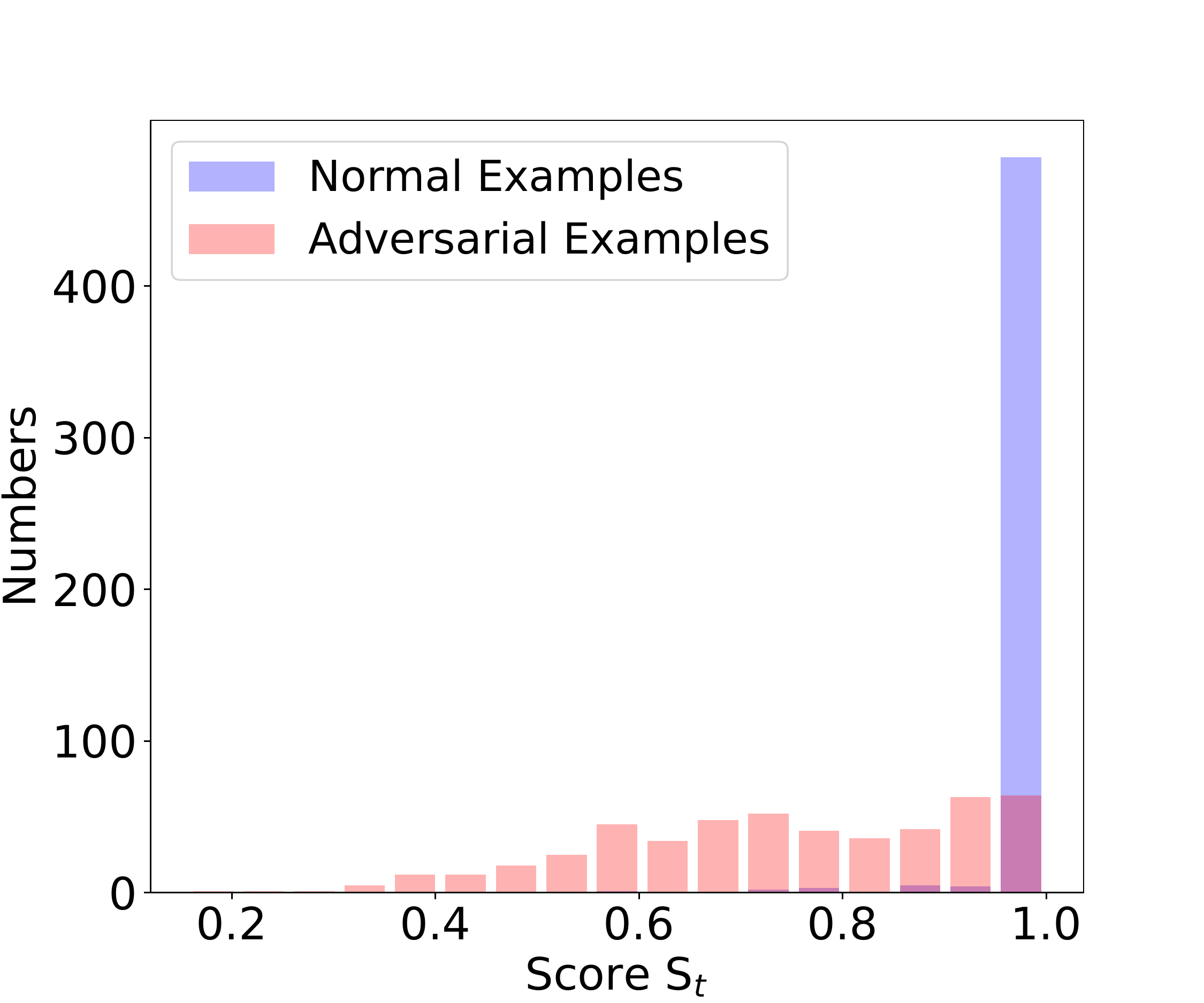}}
        \vspace{-0.2cm}
        \caption{The histogram of distinguishable scores $S_t$ defined in Eq.~(\ref{Eq:disScores}) computed for normal examples and their corresponding adversarial examples generated by attacking BERT trained on AG-NEWS with four attack methods. For adversarial examples, the manifold changes drastically after the mask and unmask operations, and therefore the classification results of the victim model are also significantly different from the original ones. In contrast, for normal examples, after the mask and unmask operations, they are still mapped back to the manifold where the normal examples are located. This makes the prediction results consistent with the original ones, and therefore their distinguishable scores $S_t$ are highly concentrated around $1.0$.}
       
        \label{fig_boxplot}
\label{Fig:STscores}
\end{figure*}
\vspace{-0.3cm}

\subsection{Attack Methods}
\label{Subsec:advAttacks}
To thoroughly evaluate the efficacy of MLMD and enable a fair comparison with other detectors  \cite{Mozes2021FrequencyGuidedWS, Mosca2022ThatIA}, we test four state-of-the-art adversarial attacks, including widely employed state-of-the-art synonym substitution-based word-level attacks and character-level attacks.

\textbf{PWWS} ({word}-level): Probability Weighted Word Saliency \cite{Ren2019GeneratingNL} is a greedy algorithm based on a synonym replacement strategy that introduces a new word replacement order determined by both the word saliency and the classification probability. The word synonym set is built via the lexical database WordNet. 

\textbf{TextFooler} ({word}-level): TextFooler \cite{Jin2020IsBR} first ranks the words in the input text by their importance and then replaces them with the most semantically similar and grammatically correct words until the prediction is altered. The substituted words are identified via cosines similarity between word embeddings of candidate tokens. 

\textbf{DeepWordBug} ({character}-level): DeepWordBug \cite{Gao2018BlackBoxGO} develops novel scoring functions to locate the most important words to be modified to make the deep classifier suffer from catastrophic prediction. Simple character-level transformations are applied to the highest-ranked words in order to minimize the edit distance of the perturbation. 

\textbf{TextBugger} ({character}-level): Similar to DeepWordBug, TextBugger \cite{Li2019TextBuggerGA} considers the more general framework of deep learning-based text understanding. It begins by identifying the most dominant token that needs to be manipulated and then chooses the optimal perturbation from a set of five generated perturbations. 

\subsection{Compared Detectors}
\label{subsec:ComparedDetectors}

We compare MLMD with two state-of-the-art adversarial detectors suggested by \cite{Mozes2021FrequencyGuidedWS, Mosca2022ThatIA}.
FGWS \cite{Mozes2021FrequencyGuidedWS} is based on the assumption that adversarial attacks prefer to replace words in the input text with low-frequency words from the training set to induce adversarial behaviors. 
It then replaces words whose frequency is below a pre-defined threshold with higher-frequency synonyms from the training data. By nature, FGWS is only suitable for word-level attacks. 

{Inspired by the use of logits-based adversarial detectors in computer vision applications, WDR \cite{Mosca2022ThatIA} first calculates the impact of a word via the word-level differential reaction (in logits) and then trains a classifier over a reaction dataset generated from both normal and adversarial examples. Unlike FGWS, WDR is suitable for detecting both word-level and character-level attacks. }

\subsection{Implementation Details}
For each dataset (3 in total as discussed in Sec.~\ref{Subsec:datasets}) and victim model architecture (4 in total as discussed in Sec.~\ref{subsec:vicmodels}) combination, we collect 1,000 examples consisting of 500 normal examples (randomly selected from the test set) and 500 associated adversarial examples created using the attack algorithms described in Sec.~\ref{Subsec:advAttacks}.
We measure the performance of FGWS \cite{Mozes2021FrequencyGuidedWS}, WDR \cite{ Mosca2022ThatIA}, and MLMD in terms of detection accuracy and F1 score.

We adapt the original implementations of FGWS and WDR \cite{Mozes2021FrequencyGuidedWS, Mosca2022ThatIA}, and fine-tune their parameters (e.g., frequency threshold of FGWS and model architecture of WDR) to obtain their best results.
For the implementation of MLMD, we used RoBERTa as the masked language model $\Phi$, unless otherwise specified. 
By default, the masking rate $r$ in Eq.~(\ref{Eq:masking}) is set to $1$ and the number of reconstructed texts $k$ in Eq.~(\ref{Eq:unMask}) is set to 3, respectively. 
We found that a simple three-layer MLP is sufficient for building the model-based classifier for our detection purposes, and although both model-based classifiers and threshold-based classifiers performed well, we choose the latter as the default method for generating detection results, unless otherwise specified. 
We report the results of the ablation study of MLMD under different mask language models and different masking rates in Sec.~\ref{Sec:ablationStudy}.

\begin{table*}[htbp]
  \centering
  \caption{Detection performance of FGWS, WDR, and MLMD on AG-NEWS and IMDB. We do not report the detection results of FGWS for the adversarial examples generated by TextBugger attack and DeepWordBug attack, since it fails to find synonyms from the training sets for some words when only characters are perturbed.}
   \resizebox{0.75\textwidth}{!}{
    \begin{tabular}{ccccccccccc}
    \toprule
    \multirow{2}[4]{*}{\textbf{Dataset}} & \multirow{2}[4]{*}{\textbf{Model}} & \multirow{2}[4]{*}{\textbf{Method}} & \multicolumn{2}{c}{\textbf{PWWS}} & \multicolumn{2}{c}{\textbf{TextFooler}} & \multicolumn{2}{c}{\textbf{TextBugger}} & \multicolumn{2}{c}{\textbf{DeepWordBug}} \\
\cmidrule{4-11}          &       & \multicolumn{1}{c}{} & \textbf{Acc.} & \textbf{F1} & \textbf{Acc.} & \textbf{F1} & \textbf{Acc.} & \textbf{F1} & \textbf{Acc.} & \textbf{F1} \\
    \midrule
    \multicolumn{1}{c}{\multirow{12}[8]{*}{AG-NEWS}} & \multirow{3}[2]{*}{BERT} & FGWS & 0.891  & 0.885  & 0.878  & 0.868  &     -    &     -    &     -    &     -    \\
          &       & WDR & \textbf{0.964 } & 0.959  & 0.970  & 0.971  & 0.937  & 0.934  & 0.907  & 0.903  \\
          &       & MLMD & 0.959  & \textbf{0.961 } & \textbf{0.983 } & \textbf{0.985 } & \textbf{0.950 } & \textbf{0.950 } & \textbf{0.938 } & \textbf{0.940 } \\
\cmidrule{2-11}    \multicolumn{1}{c}{} & \multirow{3}[2]{*}{ALBERT} & FGWS & 0.885  & 0.878  & 0.864  & 0.850  &     -    &     -    &     -    &     -    \\
          &       & WDR & 0.931  & 0.929  & 0.951  & 0.940  & 0.938  & 0.931  & 0.891  & 0.897  \\
          &       & MLMD & \textbf{0.952 } & \textbf{0.930 } & \textbf{0.984 } & \textbf{0.984 } & \textbf{0.965 } & \textbf{0.965 } & \textbf{0.943 } & \textbf{0.943 } \\
\cmidrule{2-11}          & \multirow{3}[2]{*}{CNN} & FGWS & 0.900  & 0.895  & 0.813  & 0.783  &     -    &     -    &     -    &     -    \\
          &       & WDR & 0.913  & 0.911  & 0.934  & 0.937  & 0.909  & 0.916  & 0.903  & 0.912  \\
          &       & MLMD & \textbf{0.964 } & \textbf{0.965 } & \textbf{0.971 } & \textbf{0.972 } & \textbf{0.957 } & \textbf{0.958 } & \textbf{0.956 } & \textbf{0.957 } \\
\cmidrule{2-11}          & \multirow{3}[2]{*}{LSTM} & FGWS & 0.871  & 0.862  & 0.807  & 0.776  &     -    &     -    &     -    &     -    \\
          &       & WDR & 0.910  & 0.907  & 0.932  & 0.930  & 0.893  & 0.885  & 0.904  & 0.910  \\
          &       & MLMD & \textbf{0.954 } & \textbf{0.955 } & \textbf{0.969 } & \textbf{0.967 } & \textbf{0.949 } & \textbf{0.950 } & \textbf{0.942 } & \textbf{0.946 } \\
    \midrule
    \multicolumn{1}{c}{\multirow{12}[8]{*}{IMDB}} & \multirow{3}[2]{*}{BERT} & FGWS & 0.892  & 0.881  & 0.880  & 0.867  &     -    &     -    &     -    &     -    \\
          &       & WDR & \textbf{0.946 } & \textbf{0.950 } & \textbf{0.948 } & \textbf{0.949 } & 0.950  & 0.946  & 0.915  & 0.909  \\
          &       & MLMD & 0.943  & 0.945  & 0.946  & 0.947  & \textbf{0.955 } & \textbf{0.956 } & \textbf{0.934 } & \textbf{0.936 } \\
\cmidrule{2-11}    \multicolumn{1}{c}{} & \multirow{3}[2]{*}{ALBERT} & FGWS & 0.812  & 0.839  & 0.822  & 0.814  &     -    &     -    &     -    &     -    \\
          &       & WDR & 0.801  & 0.807  & 0.873  & 0.882  & 0.882  & 0.893  & 0.799  & 0.811  \\
          &       & MLMD & \textbf{0.947 } & \textbf{0.950 } & \textbf{0.967 } & \textbf{0.970 } & \textbf{0.944 } & \textbf{0.951 } & \textbf{0.893 } & \textbf{0.895 } \\
\cmidrule{2-11}          & \multirow{3}[2]{*}{CNN} & FGWS & \textbf{0.903 } & 0.901  & 0.764  & 0.705  &     -    &     -    &     -    &     -    \\
          &       & WDR & 0.845  & 0.851  & 0.872  & 0.880  & 0.844  & 0.860  & 0.820  & 0.835  \\
          &       & MLMD & 0.898  & \textbf{0.903 } & \textbf{0.927 } & \textbf{0.929 } & \textbf{0.915 } & \textbf{0.922 } & \textbf{0.903 } & \textbf{0.908 } \\
\cmidrule{2-11}          & \multirow{3}[2]{*}{LSTM} & FGWS & 0.801  & 0.823  & 0.718  & 0.684  &     -    &     -    &     -    &     -    \\
          &       & WDR & 0.841  & 0.843  & 0.864  & 0.869  & 0.857  & 0.862  & 0.831  & 0.833  \\
          &       & MLMD & \textbf{0.886 } & \textbf{0.890 } & \textbf{0.906 } & \textbf{0.918 } & \textbf{0.900 } & \textbf{0.902 } & \textbf{0.895 } & \textbf{0.906 } \\
    \bottomrule
    \end{tabular}%
}
  \label{tab:de_re}%
\end{table*}%

\begin{table*}[htbp]
  \centering
  \caption{The detection performance of the model-based classifiers. The columns Model-C and Model denote the architecture of the adversarial classifier and the victim model, respectively. Type means the different datasets ($\Gamma$ and $\overline{\Gamma}$ in Sec.~\ref{subsubsec:ModelbasedDetector}) used for training the adversarial classifier.}
  \resizebox{0.8\textwidth}{!}{
    \begin{tabular}{cccccccccccc}
    \toprule
    \multicolumn{1}{c}{\multirow{2}[4]{*}{\textbf{Model-C}}} & \multicolumn{1}{c}{\multirow{2}[4]{*}{\textbf{Dataset}}} & \multicolumn{1}{c}{\multirow{2}[4]{*}{\textbf{Model}}} & \multirow{2}[4]{*}{\textbf{Type}} & \multicolumn{2}{c}{\textbf{PWWS}} & \multicolumn{2}{c}{\textbf{TextFooler}} & \multicolumn{2}{c}{\textbf{TextBugger}} & \multicolumn{2}{c}{\textbf{DeepWordBug}} \\
\cmidrule{5-12}          &       &       & \multicolumn{1}{c}{} & \textbf{Acc.} & \textbf{F1} & \textbf{Acc.} & \textbf{F1} & \textbf{Acc.} & \textbf{F1} & \textbf{Acc.} & \textbf{F1} \\
    \midrule
    \multicolumn{1}{c}{\multirow{8}[1]{*}{MLP}} & \multicolumn{1}{c}{\multirow{4}[1]{*}{AG-NEWS}} & \multicolumn{1}{c}{\multirow{2}[1]{*}{BERT}} & \textbf{$\Gamma$} & 0.942  & 0.937  & 0.960  & 0.955  & 0.946  & 0.940  & 0.930  & 0.930  \\
          &       &       & \textbf{$\overline{\Gamma}$} & 0.958  & 0.956  & 0.984  & 0.982  & 0.938  & 0.933  & 0.942  & 0.944  \\
          &       & \multicolumn{1}{c}{\multirow{2}[0]{*}{CNN}} & \textbf{$\Gamma$} & 0.970  & 0.969  & 0.966  & 0.965  & 0.932  & 0.931  & 0.924  & 0.921  \\
          &       &       & \textbf{$\overline{\Gamma}$} & 0.961  & 0.960  & 0.965  & 0.965  & 0.959  & 0.960  & 0.937  & 0.935  \\
          & \multicolumn{1}{c}{\multirow{4}[0]{*}{IMDB}} & \multicolumn{1}{c}{\multirow{2}[0]{*}{BERT}} & \textbf{$\Gamma$} & 0.883  & 0.866  & 0.864  & 0.837  & 0.916  & 0.902  & 0.845  & 0.823  \\
          &       &       & \textbf{$\overline{\Gamma}$} & 0.946  & 0.946  & 0.937  & 0.935  & 0.941  & 0.938  & 0.934  & 0.934  \\
          &       & \multicolumn{1}{c}{\multirow{2}[0]{*}{CNN}} & \textbf{$\Gamma$} & 0.853  & 0.851  & 0.922  & 0.925  & 0.923  & 0.924  & 0.893  & 0.893  \\
          &       &       & \textbf{$\overline{\Gamma}$} & 0.881  & 0.885  & 0.930  & 0.933  & 0.927  & 0.929  & 0.922  & 0.923  \\
          \hline
    \multicolumn{1}{c}{\multirow{8}[1]{*}{XGBoost}} & \multicolumn{1}{c}{\multirow{4}[0]{*}{AG-NEWS}} & \multicolumn{1}{c}{\multirow{2}[0]{*}{BERT}} & \textbf{$\Gamma$} & 0.944  & 0.940  & 0.956  & 0.953  & 0.935  & 0.939  & 0.937  & 0.948  \\
          &       &       & \textbf{$\overline{\Gamma}$} & 0.956  & 0.964  & 0.973  & 0.971  & 0.948  & 0.944  & 0.958  & 0.960  \\
          &       & \multicolumn{1}{c}{\multirow{2}[0]{*}{CNN}} & \textbf{$\Gamma$} & 0.960  & 0.959  & 0.963  & 0.963  & 0.950  & 0.947  & 0.933  & 0.932  \\
          &       &       & \textbf{$\overline{\Gamma}$} & 0.978  & 0.978  & 0.978  & 0.976  & 0.967  & 0.967  & 0.951  & 0.951  \\
          & \multicolumn{1}{c}{\multirow{4}[1]{*}{IMDB}} & \multicolumn{1}{c}{\multirow{2}[0]{*}{BERT}} & \textbf{$\Gamma$} & 0.929  & 0.929  & 0.934  & 0.938  & 0.937  & 0.935  & 0.927  & 0.927  \\
          &       &       & \textbf{$\overline{\Gamma}$} & 0.963  & 0.953  & 0.946  & 0.944  & 0.955  & 0.954  & 0.948  & 0.948  \\
          &       & \multicolumn{1}{c}{\multirow{2}[1]{*}{CNN}} & \textbf{$\Gamma$} & 0.857  & 0.867  & 0.935  & 0.939  & 0.931  & 0.935  & 0.898  & 0.901  \\
          &       &       & \textbf{$\overline{\Gamma}$} & 0.935  & 0.938  & 0.952  & 0.955  & 0.955  & 0.957  & 0.938  & 0.938  \\
    \bottomrule
    \end{tabular}%
}
  \label{tab:de_model}%
\end{table*}%
\vspace{-0.2cm}
\section{Experimental Results}
\subsection{Detection Performance}
\label{sec:detect_perf}
As shown in Table \ref{tab:de_re} and {Table \ref{tab:de_pre_sst2}}, compared to FGWS \cite{Mozes2021FrequencyGuidedWS} and WDR \cite{Mosca2022ThatIA}, MLMD consistently achieves comparable or better detection accuracy and F1 score across all datasets-victim model-attack method combinations ($3*4*4$ combinations). 
This validates the advantage of using a masked language model for adversarial detection. {The performance gain indicates that the changes of manifolds caused by using a masked language model and MLM objective are more suitable for detection than those induced by replacing synonyms or special tokens.} 
This observation is further supported by plotting the distinguishable score defined by Eq.~(\ref{Eq:disScores}). 
As shown in Fig.~\ref{fig_boxplot}, MLMD makes the scores of normal examples and adversarial examples obviously distinguishable.

Moreover, the results presented in both tables validate the model- and attack-agnostic nature of MLMD. When confronted with adversarial inputs engineered to attack
either Transformer-based victim models (BERT and ALBERT) or classic deep learning models (CNN and LSTM),   
MLMD performs significantly better than FGWS, exhibiting an average improvement of 1\% $\sim$ 25\% in F1 score. In addition, when compared to WDR, MLMD has shown an average increase of 1\% $\sim$ 15\% in F1 score and detection accuracy. 
When tested on the same victim model and dataset, MLMD  consistently outperforms FGWS significantly in word-level attacks. Moreover, MLMD outperforms WDR in both word- and character-level attacks by a notable margin for most cases.

An interesting observation is that FGWS typically achieves similar performance to WDR and MLMD when detecting adversarial examples generated by PWWS (best on IMDB with CNN and SST-2 with LSTM), but not for adversarial examples generated by TextFooler. 
This discrepancy may be due to the fact that FGWS identifies synonyms of infrequent words in input text via WordNet, which is the same approach used by PWWS to build a synonym set for attack.
Another noteworthy phenomenon is that perturbed texts induced by TextFooler tend to be relatively easy to identify. For instance, MLMD and WDR achieve  F1 scores of $0.985$ and $0.971$, respectively, when detecting examples generated with TextFooler on BERT trained on AG-NEWS. 
This is because TextFooler constructs adversarial examples by replacing important words with ones that have similar embeddings in a careful manner, resulting in perturbed samples residing close to the decision boundary. 
Consequently, the attack is easy to lose efficacy under masking.
Both MLMD and WDR randomly mask input words (to calculate a score that quantifies manifold change), hence invalidating adversarial examples produced by TextFooler easily.

\subsection{The Detection Performance of The Model-based Classifier}
\label{Sec:model_based_performance}
Table \ref{tab:de_model} demonstrates the performance of MLMD with model-based classifiers (i.e., a three-layer MLP and XGBoost) over AG-NEWS and IMDB {(Table \ref{tab:model_pre_sst2} for SST-2)}. 
In general, the detection performance of model-based classifiers is comparable to that of threshold-based classifiers, regardless of the model architecture. {This indicates that by performing mask and unmask procedures with the help of masked language models, normal and adversarial examples demonstrate strong distinguishable signals that can be easily separated using either thresholding or classifiers with different model architectures.}

Consistent with our expectation, regardless of the structure of the classifier, it performs better when trained on the dataset $\overline{\Gamma}$ than that on the dataset $\Gamma$, where $\overline{\Gamma}$ comprises sorted feature vectors and $\Gamma$ unsorted ones.
We speculate that the classification of original feature vectors is generally more challenging than that of sorted ones.
Moreover, the model-based classifiers only exhibit a slight advantage over the threshold-based method. As shown in Table~\ref{tab:de_model}, the advantages of the model-based classifiers are not obvious in terms of accuracy and F1 score. 
This indicates that, in our case, the confidence scores do not provide significantly more information than the labels themselves.
One possible explanation is that the mapping of the mask and the unmask operations can create a large enough gap between the normal and adversarial examples. 
Therefore, the labels alone are sufficient for distinguishing the two. As a result, there are limited improvements when using confidence scores. 
Another possibility might be the way of using confidence scores to calculate feature vectors (i.e., Eq.~\ref{Eq:featureVector}) does not fully leverage the information contained in the confidence scores, and we leave the exploration of different features as our future work.

\section{Ablation Study}
\label{Sec:ablationStudy}
We also conduct a series of ablation experiments to investigate the impact of masked language model $\Phi$ and masking rate $r$ in MLMD. 
\subsection{The Effect of Various Masked Language Models on Detection Performance}

The previous experimental results have demonstrated our proposed RoBERTa-based MLMD significantly outperforms FGWS and WDR by a wide margin, indicating the critical role played by the choice of masked language model in detection. 
Therefore, we conduct an exploratory study to further analyze how variations in the masked language model $\Phi$ affect detection results. 

Particularly, we assembled three popular masked language models (i.e., BERT, ALBERT, and RoBERTa) in MLMD, resulting in detectors for SST-2, AG-NEWS, and IMDB. For each dataset, such as SST-2, all masked language models are compared on a balanced set, which is constructed of normal examples and corresponding adversarial examples generated via TextFooler attacks on four victim models trained on the dataset. 

As shown in Table \ref{tab:vaious_effect}, all masked language models achieve comparable performance, outperforming FGWS and WDR in most cases in terms of accuracy and F1 score. This suggests that MLMD does not depend on a specific masked language model, and general masked language models possess the capability to distinguish adversarial examples in the MLMD framework. However, one of the most compelling signs of the results in Table \ref{tab:vaious_effect} is that by varying the masked language model component from BERT to RoBERTa, the detection performance exhibits a slight upward trend. One of the most likely reasons is that they have different approximation capacities when fitting the manifold of normal examples. We attribute this phenomenon to the following aspects. 

On the one hand, analyzing these three masked language models from the perspective of pre-training data, BERT and ALBERT were pre-trained on a concatenation of English Wikipedia and Bookcorpus. However, RoBERTa was trained on a combination of five large-scale datasets, including Bookcorpus, English Wikipedia, CC-NEWS, Openwebtext, and Stories {\cite{Liu2019RoBERTaAR}}. The training data of RoBERTa is not only significantly larger than that of BERT but also more heterogeneous. As a result,  RoBERTa is capable of capturing more comprehensive and accurate language representations of normal examples, which allows it to better approximate the manifold of normal examples than the other two models.

On the other hand, the self-supervised training objectives of the three models are designed differently. 
For the MLM objective, BERT uniformly selects 15\% of the input tokens for possible replacement. Among the selected tokens, 80\% are replaced with the [MASK] token, 10\% are unchanged, and 10\% are replaced by a randomly selected token. 
BERT performed only one masking operation during data preprocessing, resulting in a static masking strategy. 
In contrast, ALBERT generates masked texts using n-gram masking with the length of each n-gram mask selected randomly, which means a masked position can consist of up to an n-gram of complete words. 
Since an n-gram contains more information than a single token, 
ALBERT can better model the input manifold and obtain the complete semantics of the words that make up the sentence. 
However, RoBERTa further improves BERT by adopting a dynamic masking strategy that duplicates training data 10 times so that each text is masked in 10 different ways over the 40 epochs of pre-training. This enables an example to have different masked copies, which improves the randomness of the input text and the learning ability of the model. Through dynamic masking, RoBERTa can better focus on mining the intrinsic information of the manifold of normal examples.

Finally, the three models also differ in their training techniques. In short, RoBERTa was trained for a longer period with large batches, resulting in improved perplexity for the masked language modeling objective. 
 
To summarize, the manifold of normal examples approximated by RoBERTa is closer to the underlying manifold of normal data, resulting in improved performance of MLMD.

\begin{table}[thbp]
  \centering
  \caption{The effect of various masked language models on detection results. We consider the three most popular masked language models: BERT, ALBERT, and RoBERTa.}
  \resizebox{0.485\textwidth}{!}{
    \begin{tabular}{cccccccc}
    \toprule
    \multirow{2}[4]{*}{\textbf{Dataset}} & \multirow{2}[4]{*}{\textbf{Model}} & \multicolumn{2}{c}{\textbf{BERT}} & \multicolumn{2}{c}{\textbf{ALBERT}} & \multicolumn{2}{c}{\textbf{RoBERTa}} \\
\cmidrule{3-8}          & \multicolumn{1}{c}{} & \textbf{Acc.} & \textbf{F1} & \textbf{Acc.} & \textbf{F1} & \textbf{Acc.} & \textbf{F1} \\
   \midrule
    \multirow{4}[2]{*}{\makecell{AG-\\NEWS}} & BERT & 0.961 & 0.961 & 0.961 & 0.961  & \textbf{0.983 } & \textbf{0.985 } \\
          & ALBERT & 0.971 & 0.971 & 0.977 & 0.977  & \textbf{0.984 } & \textbf{0.986 } \\
          & CNN & 0.965 & 0.966 & 0.968 & 0.969  & \textbf{0.971 } & \textbf{0.972 } \\
          & LSTM & 0.968 & 0.969 & \textbf{0.969} & \textbf{0.969 } & 0.969  & 0.967  \\
    \midrule
    \multirow{4}[2]{*}{IMDB} & BERT & 0.941 & 0.942 & 0.944 & 0.944  & \textbf{0.946 } & \textbf{0.947 } \\
          & ALBERT & 0.923 & 0.924 & 0.944 & 0.944  & \textbf{0.967 } & \textbf{0.970 } \\
          & CNN & 0.912 & 0.913 & 0.922 & 0.924  & \textbf{0.927 } & \textbf{0.929 } \\
          & LSTM & 0.901 & 0.905 & 0.899 & 0.902  & \textbf{0.906 } & \textbf{0.908 } \\
    \midrule
    \multirow{4}[2]{*}{SST-2} & BERT & 0.839 & 0.846 & 0.847 & 0.853  & \textbf{0.901 } & \textbf{0.903 } \\
          & ALBERT & 0.835 & 0.844 & 0.842 & 0.848  & \textbf{0.850 } & \textbf{0.857 } \\
          & CNN & 0.831 & 0.844 & 0.833 & 0.850  & \textbf{0.851 } & \textbf{0.860 } \\
          & LSTM & 0.819 & 0.831 & 0.819 & 0.832  & \textbf{0.832 } & \textbf{0.843 } \\
    \bottomrule
    \end{tabular}%
}
  \label{tab:vaious_effect}%
\end{table}%

\subsection{The Effect of Removing the Masked Language Model Component on Detection Performance}
MLMD variants built by different masked language models show quite competitive performance under the same combination of datasets and victim models, which further indicates that the pre-trained masked language model has the ability to approximate the manifold of normal examples and also implies that the effectiveness of MLMD is not reliant on a specific $\Phi$. In this section, we further consider a more rigorous setting, namely, directly removing $\Phi$ from MLMD (we call this variant MLMD-M) to study the importance of the masked language model in detecting adversarial examples. 

{We conducted experiments on the AG-NEWS dataset, and the results are reported in Table \ref{tab:mask_effect}. We notice that the removal of the masked language model will  impair the detection performance with an average loss of 1\% $\sim$ 5\% in detection accuracy and F1 score. That said, MLMD-M is comparable to the performance of WDR, whose performance is discussed in Sec.~\ref{sec:detect_perf}. This once again agrees with our analyses in Sec.~\ref{subsec:ComparedDetectors}, WDR differs from MLMD in two aspects: it measures the differences in logits and it does not have an unmask procedure. Since the softmax layer is parameter-free, measuring logits and measuring confidence scores are equivalent. As such, this observation confirms that the masked language model can project corrupted examples back to the manifold of normal data, which is the rationale of MLMD.}


\begin{table*}[htbp]
  \centering
  \caption{The effect of removing the masked language model component on detection performance. MLMD-M represents that only the masked texts generated by the mask procedure are used for extracting distinguishable signals, rather than reconstructed texts generated by using the unmask procedure.}
    \begin{tabular}{cp{4.8em}cccccccc}
    \toprule
    \multirow{2}[4]{*}{\textbf{Model}} & \multicolumn{1}{c}{\multirow{2}[4]{*}{\textbf{Method}}} & \multicolumn{2}{c}{\textbf{PWWS}} & \multicolumn{2}{c}{\textbf{TextFooler}} & \multicolumn{2}{c}{\textbf{TextBugger}} & \multicolumn{2}{c}{\textbf{DeepWordBug}} \\
\cmidrule{3-10}          & \multicolumn{1}{c}{} & \textbf{Acc.} & \textbf{F1} & \textbf{Acc.} & \textbf{F1} & \textbf{Acc.} & \textbf{F1} & \textbf{Acc.} & \textbf{F1} \\
    \midrule
    \multirow{2}[2]{*}{BERT} & 
    \multicolumn{1}{c}{MLMD-M} & 0.931 & 0.931 & 0.961 & 0.962 & 0.925 & 0.926 & 0.893 & 0.895 \\
          &\multicolumn{1}{c}{MLMD} & \textbf{0.959} & \textbf{0.961} & \textbf{0.983} & \textbf{0.985} & \textbf{0.950} & \textbf{0.950} & \textbf{0.938} & \textbf{0.950} \\
    \midrule
    \multirow{2}[2]{*}{ALBERT} & \multicolumn{1}{c}{MLMD-M} & 0.942 & \textbf{0.945 } & 0.951 & 0.953 & 0.945 & 0.947 & 0.930 & 0.934 \\
          & \multicolumn{1}{c}{MLMD} & \textbf{0.952} & 0.930 & \textbf{0.984} & \textbf{0.984} & \textbf{0.965} & \textbf{0.965} & \textbf{0.943} & \textbf{0.943} \\
    \midrule
    \multirow{2}[2]{*}{CNN} & \multicolumn{1}{c}{MLMD-M} & 0.938 & 0.939 & 0.941 & 0.939 & 0.936 & 0.939 & 0.906 & 0.904 \\
          & \multicolumn{1}{c}{MLMD} & \textbf{0.964} & \textbf{0.965} & \textbf{0.971} & \textbf{0.972} & \textbf{0.957} & \textbf{0.958} & \textbf{0.956} & \textbf{0.957} \\
    \midrule
    \multirow{2}[2]{*}{LSTM} & \multicolumn{1}{c}{MLMD-M} & 0.910 & 0.911 & 0.943 & 0.944 & 0.920 & 0.922 & 0.896 & 0.895 \\
          & \multicolumn{1}{c}{MLMD} & \textbf{0.954} & \textbf{0.955} & \textbf{0.969} & \textbf{0.967} & \textbf{0.949} & \textbf{0.950} & \textbf{0.942} & \textbf{0.946} \\
    \bottomrule
    \end{tabular}%
  \label{tab:mask_effect}%
\end{table*}%

\subsection{The Effect of Masking Rate on Detection Performance}
We now explore the impact of the masking rate of the mask operation on the detection performance of MLMD. We conduct experiments with the combination of BERT, TextFooler, and AG-NEWS. 
We create a series of masked texts by randomly selecting some words to mask in the input text according to a rate $r$, and we adjust the value of $r$ from 10\% to 100\% to get the detection results.

\begin{figure}[t]
    \centering 
    \includegraphics[height=5.2cm]{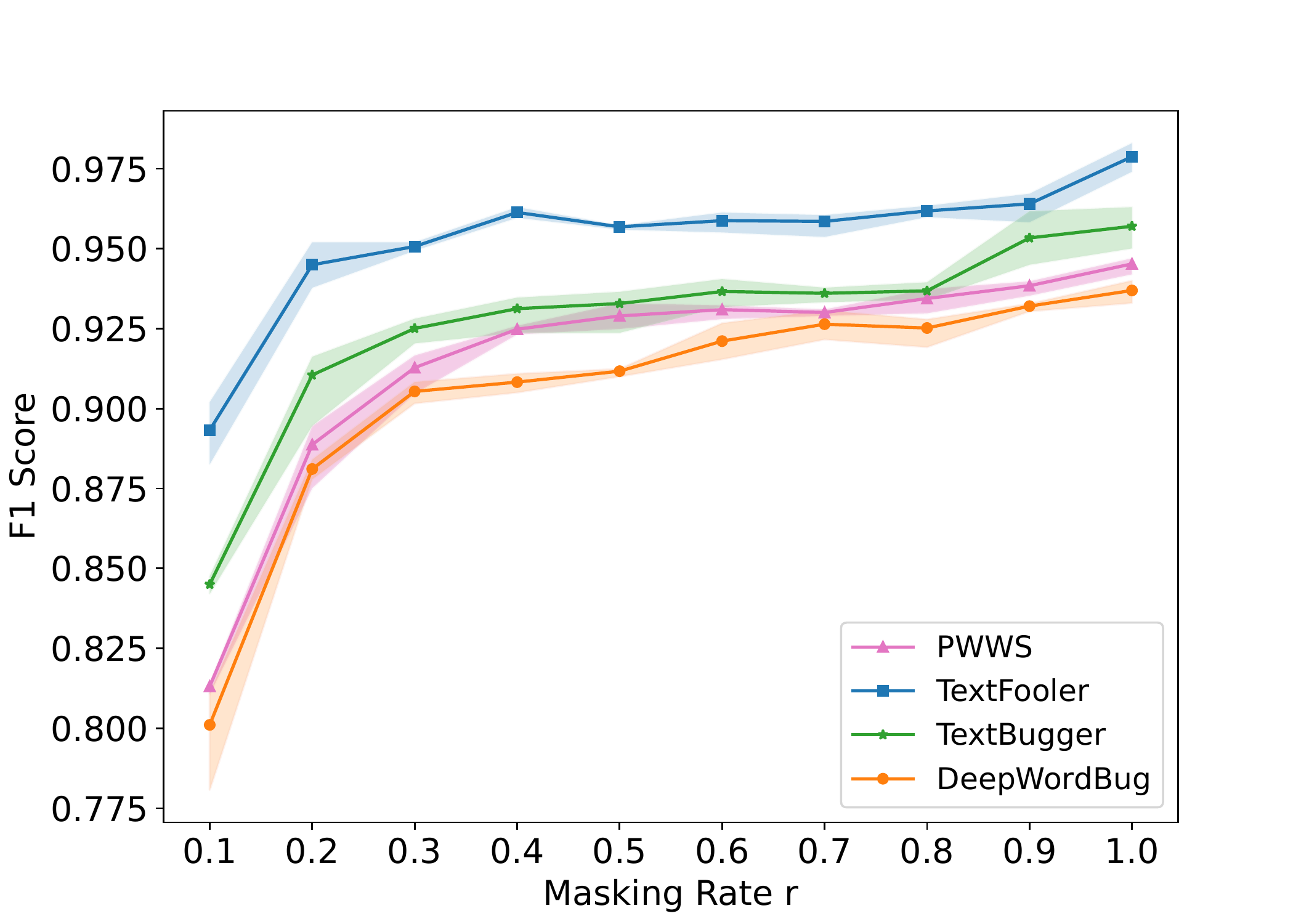}
    \caption{The effect of masking rate on detection performance. Results averaged over {3} runs.}
    \label{Fig:masking_rate}

\end{figure}

From Fig.~\ref{Fig:masking_rate}, we can see an optimal value for the masking rate across four attack methods. Increasing the masking rate results in improved detection performance. The higher the masking rate, the greater the probability that the critical words in the adversarial example will be masked and unmasked. With a larger masking rate, we have a higher probability of collecting distinguishable information about the manifold changes, which can provide strong evidence for detection.

\subsection{Fine-tuning Masked Language Models for Detection}
Gururangan et al. reported that performing the MLM objective on the target domain with unlabeled data can also help to improve downstream task performance \cite{Gururangan2020DontSP}. 
{This finding motivates us to fine-tune the masked language model $\Phi$ to further improve the detection performance of MLMD. Fixing the other settings, we fine-tune the RoBERTa component of MLMD on the target domain (i.e., AG-NEWS and IMDB), which we refer to as MLMD-F. Due to limited computing power, we tune RoBERTa for 20 epochs (as opposed to the original 40) with a reduced batch size (128 v.s. 8000).} 

The experimental results presented in Table \ref{tab:de_fine} show that MLMD-F has a general growth trend compared with the original MLMD in terms of accuracy and F1 score.  
After fine-tuning, the manifold approximated by RoBERTa is more matched with the manifold of examples in the target domain, resulting in improved detection performance compared with the original RoBERTa-based MLMD. 
This suggests that fine-tuning can be an effective method to further improve performance.

However, some exceptions are observed in Table~\ref{tab:de_fine}.
For example, when detecting adversarial examples crafted by TextFooler against BERT trained on AG-NEWS, MLMD-F results in a decline in detection accuracy and F1 scores. This phenomenon could be due to the fact that the original MLMD already has a strong detection performance on TextFooler, leaving little room for improvement, or the RoBERTa is not sufficiently fine-tuned. 
It is also worth noting that although the detection results of the MLMD-F are improved compared to the original MLMD, the gain is not significant (i.e., {less than 1\% on average}). 
This is likely due to the limited computing power available for fine-tuning, resulting in RoBERTa's imperfect approximation of the data manifold on the target domain. 
Therefore, the detection performance still heavily relies on the manifold initially fitted by the masked language model.

\begin{table*}[htbp]
  \centering
  \caption{
  Detection results of the fine-tuned masked language model as MLMD's component. MLMD-F denotes the MLMD detection method instantiated by the fine-tuned RoBERTa. 
  }
  \resizebox{0.78\textwidth}{!}{
    \begin{tabular}{ccccccccccc}
    \toprule
    \multirow{2}[4]{*}{\textbf{Dataset}} & \multirow{2}[4]{*}{\textbf{Model}} & \multicolumn{1}{c}{\multirow{2}[4]{*}{\textbf{Method}}} & \multicolumn{2}{c}{\textbf{PWWS}} & \multicolumn{2}{c}{\textbf{TextFooler}} & \multicolumn{2}{c}{\textbf{TextBugger}} & \multicolumn{2}{c}{\textbf{DeepWordBug}} \\
\cmidrule{4-11}          &       &       & \textbf{Acc.} & \textbf{F1} & \textbf{Acc.} & \textbf{F1} & \textbf{Acc.} & \textbf{F1} & \textbf{Acc.} & \textbf{F1} \\
    \midrule
    \multicolumn{1}{c}{\multirow{8}[8]{*}{AG-NEWS}} & \multirow{2}[2]{*}{BERT} & \multicolumn{1}{c}{MLMD} & 0.959 & 0.961 & \textbf{0.983} & \textbf{0.985} & 0.950 & 0.950 & 0.938 & 0.940 \\
          &       & MLMD-F & \textbf{0.962} & \textbf{0.964} & 0.981 & 0.982 & \textbf{0.956} & \textbf{0.959} & \textbf{0.943} & \textbf{0.943} \\
\cmidrule{2-11}          & \multirow{2}[2]{*}{ALBERT} & \multicolumn{1}{c}{MLMD} & 0.952 & 0.930 & 0.984 & 0.984 & 0.965 & 0.965 & 0.943 & 0.943 \\
          &       & MLMD-F & \textbf{0.970} & \textbf{0.971} & \textbf{0.984} & \textbf{0.987} & \textbf{0.973} & \textbf{0.974} & \textbf{0.966} & \textbf{0.967} \\
\cmidrule{2-11}          & \multirow{2}[2]{*}{CNN} & \multicolumn{1}{c}{MLMD} & 0.964 & 0.965 & 0.971 & 0.972 & 0.957 & 0.958 & 0.956 & 0.957 \\
          &       & MLMD-F & \textbf{0.969} & \textbf{0.971} & \textbf{0.987} & \textbf{0.986} & \textbf{0.972} & \textbf{0.975} & \textbf{0.973} & \textbf{0.973} \\
\cmidrule{2-11}          & \multirow{2}[2]{*}{LSTM} & \multicolumn{1}{c}{MLMD} & 0.954 & 0.955 & 0.969 & 0.967 & 0.949 & 0.950 & 0.942 & 0.946 \\
          &       & MLMD-F & \textbf{0.973} & \textbf{0.974} & \textbf{0.976} & \textbf{0.976} & \textbf{0.969} & \textbf{0.970} & \textbf{0.959} & \textbf{0.964} \\
    \midrule
    \multicolumn{1}{c}{\multirow{8}[8]{*}{IMDB}} & \multirow{2}[2]{*}{BERT} & \multicolumn{1}{c}{MLMD} & 0.943 & 0.945 & 0.946 & 0.947 & 0.955 & 0.956 & 0.934 & 0.936 \\
          &       & MLMD-F & \textbf{0.959} & \textbf{0.970} & \textbf{0.952} & \textbf{0.953} & \textbf{0.956} & \textbf{0.957} & \textbf{0.949} & \textbf{0.950} \\
\cmidrule{2-11}          & \multirow{2}[2]{*}{ALBERT} & \multicolumn{1}{c}{MLMD} & \textbf{0.947 } & 0.950 & \textbf{0.967} & \textbf{0.970} & 0.944 & 0.951 & 0.893 & 0.895 \\
          &       & MLMD-F & 0.945 & \textbf{0.950} & 0.967 & 0.968 & \textbf{0.955} & \textbf{0.959} & \textbf{0.923} & \textbf{0.925} \\
\cmidrule{2-11}          & \multirow{2}[2]{*}{CNN} & \multicolumn{1}{c}{MLMD} & \textbf{0.898} & \textbf{0.903} & 0.927 & 0.929 & 0.915 & 0.922 & 0.903 & 0.908 \\
          &       & MLMD-F & 0.893 & 0.896 & \textbf{0.941} & \textbf{0.942} & \textbf{0.935} & \textbf{0.937} & \textbf{0.911} & \textbf{0.914} \\
\cmidrule{2-11}          & \multirow{2}[2]{*}{LSTM} & \multicolumn{1}{c}{MLMD} & 0.886 & 0.890 & 0.906 & 0.918 & 0.900 & 0.902 & 0.895 & 0.906 \\
          &       & MLMD-F & \textbf{0.921} & \textbf{0.919} & \textbf{0.914} & \textbf{0.916} & \textbf{0.927} & \textbf{0.932} & \textbf{0.905} & \textbf{0.906} \\
    \bottomrule
    \end{tabular}%
}
  \label{tab:de_fine}%
\end{table*}%

\section{Discussions} 
As an online detector, although MLMD achieves impressive performance, it comes at the cost of delay (i.e., time complexity) and extra power (i.e., computation complexity).
In particular, MLMD requires a feature vector of length $r*n*k$ for each textual example (with length $n$), resulting in $r*n*k$ calls to the victim $f$ during online inference stage. The complexity is dominated by $n$ as $r$ is typically 1 and $k$ is normally small ($k=3$ as shown above). Note that the complexity of MLMD is at the same order as WDR (WDR \cite{Mosca2022ThatIA} computes logits reaction with word-by-word masking). In contrast, the complexity of FGWS is lower as it only replaces words with frequencies lower than a threshold \cite{Mozes2021FrequencyGuidedWS}. 

{Clearly, the online complexity of MLMD (and also WDR) can be mitigated by deploying multiple copies of the victim $f$ and calling them in parallel}. 
Additionally, the online complexity can be reduced by selecting words for masking based on appropriate criteria (similar to the idea of \cite{Mozes2021FrequencyGuidedWS}, which effectively makes $r<1$), {as well as (dynamically) masking a few words at one time.}
We leave the study of these possible improvements for our future work.

Moreover, attackers can adapt their strategy to make stronger attacks once they become aware of the deployed defense \cite{Carlini2017AdversarialEA, Zhang2021SelfSupervisedAE}. In particular, adaptive attackers need to optimize an adversarial perturbation that simultaneously maximizes the loss of the victim $f$ while evading the MLMD detector. As MLMD captures manifold changes for feature engineering, this is equivalent to finding adversarial examples that do not induce manifold change through the mask and unmask procedures.  However, the masked language model only {fits the manifold of normal data \cite{Ng2020SSMBASM, Hendrycks2020PretrainedTI}}, which contradicts the objective of the attacker. As such, depending on how well the masked language model fits the manifold (of normal data), 
MLMD can significantly increase the cost of adaptive attacks or even make them fail.

\section{Conclusion}
Based on the fact that the masked language model can fit the manifold of normal examples and the off-manifold conjecture of adversarial examples, this paper proposes a novel textual adversarial example detector MLMD.
It is built on top of the difference in manifold change when a suspicious example passes through the mask and unmask procedures. Extensive experiments demonstrate that MLMD possesses remarkable generalization capabilities across different victim models, datasets, and attacks, outperforming other SOTA detectors. As a top-up \textit{plug and play} module that works with already-deployed victim machine learning systems, MLMD is a useful tool that complements existing adversarial defenses. 
Taking full advantage of the masked language model to improve adversarial robustness is an interesting problem that requires further in-depth investigation.

\begin{acks}
This work is supported by the Technology Innovation and Application Development of Chongqing Science and Technology Commission (CSTC2019JSCX-KJFP-X0022), and by the Open Fund of Science and Technology on Parallel and Distributed Processing Laboratory (WDZC20205250114).
\end{acks}

\bibliographystyle{ACM-Reference-Format}
\bibliography{ref1}
\clearpage

\appendix

\begin{table*}[hb]
  \centering
  \caption{Detection performance of FGWS, WDR, and MLMD on SST-2. We do not report the detection results of FGWS for the adversarial examples generated by TextBugger attack and DeepWordBug attack, since it fails to find synonyms from the training sets for some words when only characters are perturbed.}
  \resizebox{0.75\textwidth}{!}{
    \begin{tabular}{ccccccccccc}
    \toprule
    \multirow{2}[2]{*}{\textbf{Dataset}} & \multirow{2}[2]{*}{\textbf{Model}} & \multirow{2}[2]{*}{\textbf{Method}} & \multicolumn{2}{c}{\textbf{PWWS}} & \multicolumn{2}{c}{\textbf{TextFooler}} & \multicolumn{2}{c}{\textbf{TextBugger}} & \multicolumn{2}{c}{\textbf{DeepWordBug}} \\
    \cmidrule{4-11}&       & \multicolumn{1}{c}{} & \textbf{Acc.} & \textbf{F1} & \textbf{Acc.} & \textbf{F1} & \textbf{Acc.} & \textbf{F1} & \textbf{Acc.} & \textbf{F1} \\
    \midrule
    \multicolumn{1}{c}{\multirow{12}[8]{*}{SST-2}} & \multirow{3}[2]{*}{BERT} & FGWS & 0.821 & 0.790  & 0.769 & 0.711 &     -    &     -    &     -    &     -    \\
          &       & WDR & 0.787 & 0.800   & 0.790  & 0.807 & 0.783 & 0.807 & 0.728 & 0.737 \\
          &       & MLMD & \textbf{0.837} & \textbf{0.848} & \textbf{0.901} & \textbf{0.903} & \textbf{0.852} & \textbf{0.855} & \textbf{0.830} & \textbf{0.839} \\
\cmidrule{2-11}    \multicolumn{1}{c}{} & \multirow{3}[2]{*}{ALBERT} & FGWS & 0.811 & 0.779 & 0.738 & 0.663 &     -    &     -    &     -    &      -     \\
          &       & WDR & 0.730  & 0.757 & 0.747 & 0.773 & 0.714 & 0.755 & 0.674 & 0.698 \\
          &       & MLMD & \textbf{0.837} & \textbf{0.848} & \textbf{0.850} & \textbf{0.857} & \textbf{0.849} & \textbf{0.855} & \textbf{0.826} & \textbf{0.836} \\
\cmidrule{2-11}          & \multirow{3}[2]{*}{CNN} & FGWS & 0.798 & 0.767 & 0.689 & 0.588 &     -    &     -    &     -    &     -    \\
          &       & WDR & 0.715 & 0.769 & 0.719 & 0.773 & 0.703 & 0.768 & 0.721 & 0.776 \\
          &       & MLMD & \textbf{0.811} & \textbf{0.829} & \textbf{0.851} & \textbf{0.860} & \textbf{0.842} & \textbf{0.850} & \textbf{0.817} & \textbf{0.831} \\
\cmidrule{2-11}          & \multirow{3}[2]{*}{LSTM} & FGWS & \textbf{0.801} & \textbf{0.811} & 0.678 & 0.566 &     -    &     -    &     -    &     -    \\
          &       & WDR & 0.736 & 0.776 & 0.751 & 0.788 & 0.724 & 0.778 & 0.686 & 0.742 \\
          &       & MLMD & 0.778 & 0.809 & \textbf{0.832} & \textbf{0.843} & \textbf{0.832} & \textbf{0.843} & \textbf{0.788} & \textbf{0.805} \\
    \bottomrule
    \end{tabular}%
}
  \label{tab:de_pre_sst2}%
\end{table*}%
\par
\begin{table*}[hb]
  \centering
  \caption{The detection performance of the model-based classifiers on SST-2. The columns Model-C and Model denote the architecture of the adversarial classifier and the victim model, respectively. Type means the different datasets ($\Gamma$ and $\overline{\Gamma}$ in Sec.~\ref{subsubsec:ModelbasedDetector}) used for training the adversarial classifier.}
   \resizebox{0.75\textwidth}{!}{
    \begin{tabular}{ccccccccccc}
    \toprule
    \multicolumn{1}{c}{\multirow{2}[4]{*}{\textbf{Model-C}}} & \multicolumn{1}{c}{\multirow{2}[4]{*}{\textbf{Model}}} & \multirow{2}[4]{*}{\textbf{Type}} & \multicolumn{2}{c}{\textbf{PWWS}} & \multicolumn{2}{c}{\textbf{TextFooler}} & \multicolumn{2}{c}{\textbf{TextBugger}} & \multicolumn{2}{c}{\textbf{DeepWordBug}} \\
\cmidrule{4-11}          &       & \multicolumn{1}{c}{} & \textbf{Acc.} & \textbf{F1.} & \textbf{Acc.} & \textbf{F1.} & \textbf{Acc.} & \textbf{F1.} & \textbf{Acc.} & \textbf{F1.} \\
    \midrule
    \multicolumn{1}{c}{\multirow{4}[2]{*}{MLP}} & \multicolumn{1}{c}{\multirow{2}[1]{*}{BERT}} & \textbf{$\Gamma$} & 0.813  & 0.820  & 0.824  & 0.815  & 0.846  & 0.837  & 0.827  & 0.819  \\
          &       & \textbf{$\overline{\Gamma}$} & 0.831  & 0.824  & 0.844  & 0.842  & 0.860  & 0.855  & 0.815  & 0.801  \\
          & \multicolumn{1}{c}{\multirow{2}[1]{*}{CNN}} & \textbf{$\Gamma$} & 0.809  & 0.814  & 0.829  & 0.824  & 0.808  & 0.799  & 0.789  & 0.799  \\
          &       & \textbf{$\overline{\Gamma}$} & 0.809  & 0.811  & 0.854  & 0.861  & 0.844  & 0.848  & 0.801  & 0.789  \\
    \midrule
    \multicolumn{1}{c}{\multirow{4}[2]{*}{XGBoost}} & \multicolumn{1}{c}{\multirow{2}[1]{*}{BERT}} & \textbf{$\Gamma$} & 0.831  & 0.829  & 0.886  & 0.882  & 0.877  & 0.897  & 0.853  & 0.845  \\
          &       & \textbf{$\overline{\Gamma}$} & 0.837  & 0.833  & 0.889  & 0.883  & 0.887  & 0.889  & 0.863  & 0.866  \\
          & \multicolumn{1}{c}{\multirow{2}[1]{*}{CNN}} & \textbf{$\Gamma$} & 0.827  & 0.829  & 0.854  & 0.866  & 0.843  & 0.853  & 0.827  & 0.828  \\
          &       & \textbf{$\overline{\Gamma}$} & 0.859  & 0.860  & 0.873  & 0.884  & 0.859  & 0.866  & 0.841  & 0.843  \\
    \bottomrule
    \end{tabular}%
}
  \label{tab:model_pre_sst2}%
\end{table*}%

\section*{Appendix} 
The Appendix provides detection results for the threshold-based classifier and the model-based classifiers on the SST-2 dataset.

\subsection*{The Detection Performance of The Threshold-based Classifier}
The detection performance results of FGWS, WDR, and MLMD on SST-2 are shown in Table \ref{tab:de_pre_sst2}. Consistent with the results on the IMDB and AG-NEWS datasets, our method outperforms FGWS and WDR in terms of accuracy and F1 score. Note that the average length of examples in the SST dataset is only 20 words, which is too short for FGWS to provide stable results on different attacks (PWWS v.s. TextFooler). This is because FGWS only replaces a few words of the input based on frequency, and frequency is not a stable measurement for short inputs. 
\vspace{1cm}
\subsection*{The Detection Performance of The Model-based Classifier}

Table \ref{tab:model_pre_sst2} lists the detection performances of different model-based classifiers on the SST-2 dataset. It is not difficult to find that the model-based classifiers and the threshold-based classifier have similar detection effects. Additionally, as expected, classifiers trained on $\overline{\Gamma}$ are more effective because they use additional information about the feature vector (i.e., the order of its elements).
\end{document}